\documentclass[11pt]{article}
\usepackage{blois}

\usepackage[a4paper,top=1.9cm,right=1.9cm,left=1.9cm,bottom=3.67cm]{geometry}

\usepackage{graphicx}
\usepackage{subfigure}
\usepackage{lineno}
\usepackage{url}

\def\be{\begin{equation}}
\def\ee{\end{equation}}
\def\bea{\begin{eqnarray}}
\def\eea{\end{eqnarray}}

\begin{document}
\vspace*{4cm}
\title{Results on QCD jet production at ATLAS and CMS}

\author{ Christopher J. Meyer, on behalf of the ATLAS and CMS Collaborations.}

\address{The Enrico Fermi Institute, The University of Chicago,\\
5640 South Ellis Avenue, Chicago, IL 60637.}

\maketitle\abstracts{
The production of jets at the Large Hadron Collider (LHC) at $\sqrt{s}=7$~TeV is summarized, including results from both the ATLAS and CMS detectors.
Current jet performance is described, followed by inclusive jet and multi-jet measurements in various final state configurations. 
Finally some results on heavy flavour and jet substructure are presented.
}

At both the ATLAS \cite{ATLAS:det} and CMS \cite{CMS:det} detectors in the LHC, jets serve as a proxy to final state partons.
Following the hard collision they undergo parton showering, hadronisation, and subsequently interact in the surrounding detector.
To reconstruct and calibrate the constituents of jets, CMS uses a particle flow method which employs several subdetectors \cite{Chatrchyan:2011je} while ATLAS forms jets using finely segmented calorimeters.\cite{Aad:2011he}
In both cases a pileup offset correction is applied to remove additional energy from multiple proton-proton collisions.

The anti-$k_{T}$ clustering algorithm \cite{Cacciari:2008gp} is the preferred choice for jet reconstruction, with other methods such as Cambridge-Aachen \cite{Dokshitzer:1997in} used for jet substructure.
For cross section measurements a radius between $0.4 \le R \le 0.7$ is used, while larger jet radii ($R \ge 1.0$) are used for jet substructure studies.

\begin{figure}
\centering
\subfigure{
   \includegraphics[height=5.0cm]{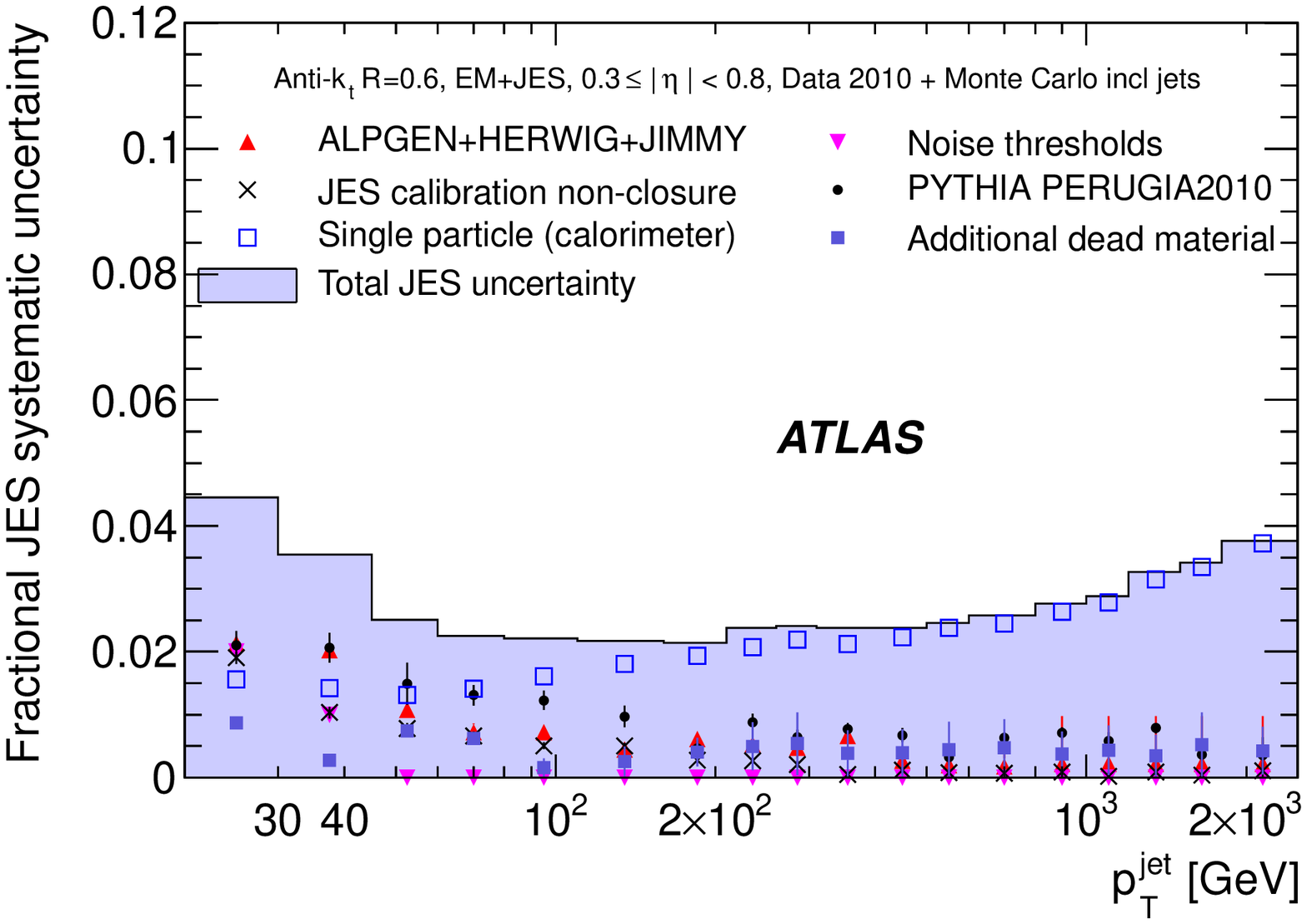}
}
\subfigure{
   \includegraphics[height=5.0cm]{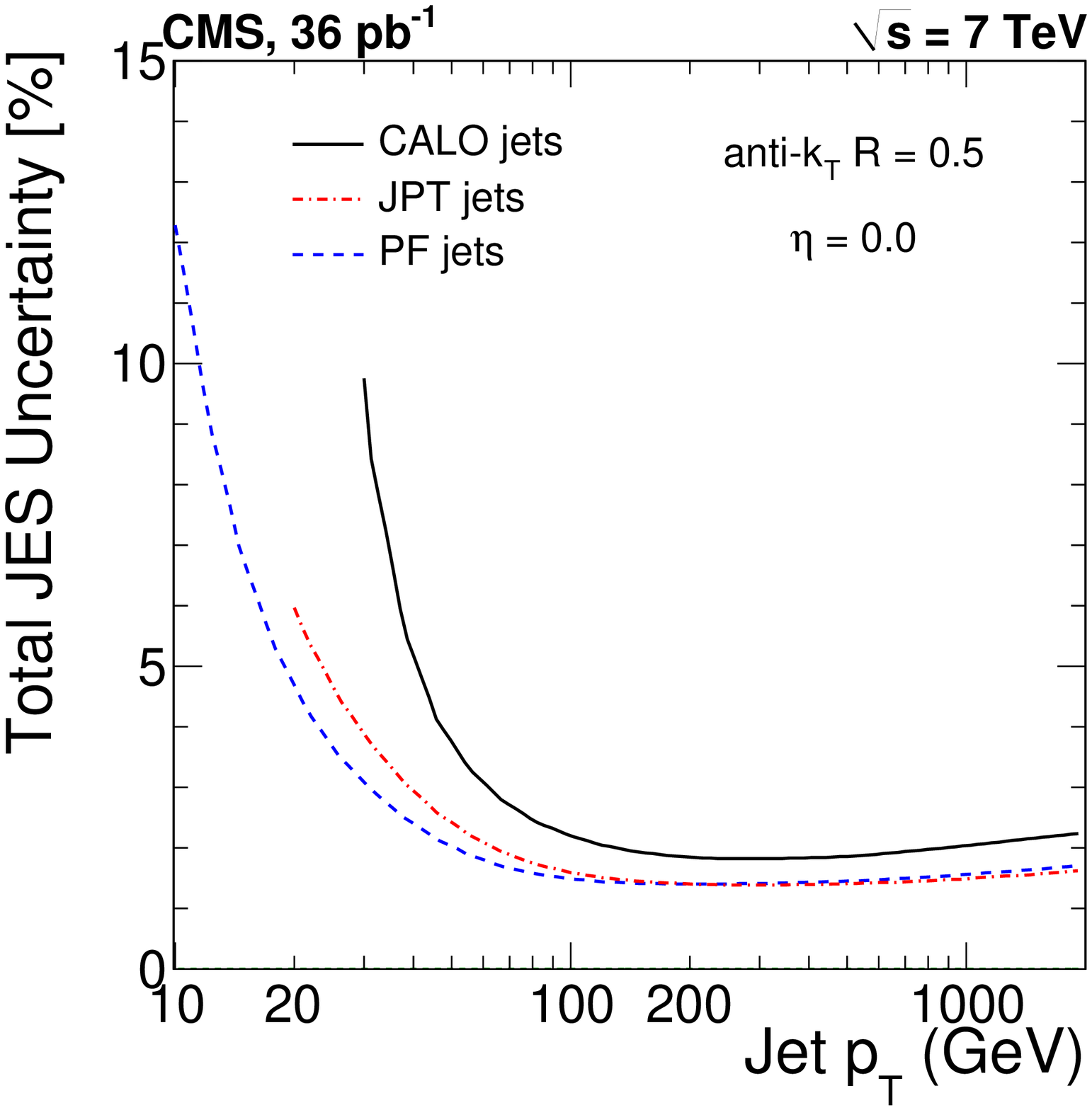}
}
\caption{Fractional uncertainty on the jet energy calibration as a function of jet $p_{T}$ in ATLAS \protect\cite{Chatrchyan:2011je} and CMS.\protect\cite{Aad:2011he}}
\label{fig:JESunc}
\end{figure}

The 2010 jet energy calibration is derived using Monte Carlo (MC) simulation tuned using test beam and early collision data.
The response is derived by comparing fully simulated and reconstructed jets to truth jets, with in situ techniques such as multijet, photon-jet, and Z-jet balance used as cross checks in data.
The uncertainty on the derived jet energy calibration, shown in Figure \ref{fig:JESunc}, is often the dominant source of experimental error on cross section measurements.

\begin{figure}
\centering
\subfigure{
   \includegraphics[height=6.5cm]{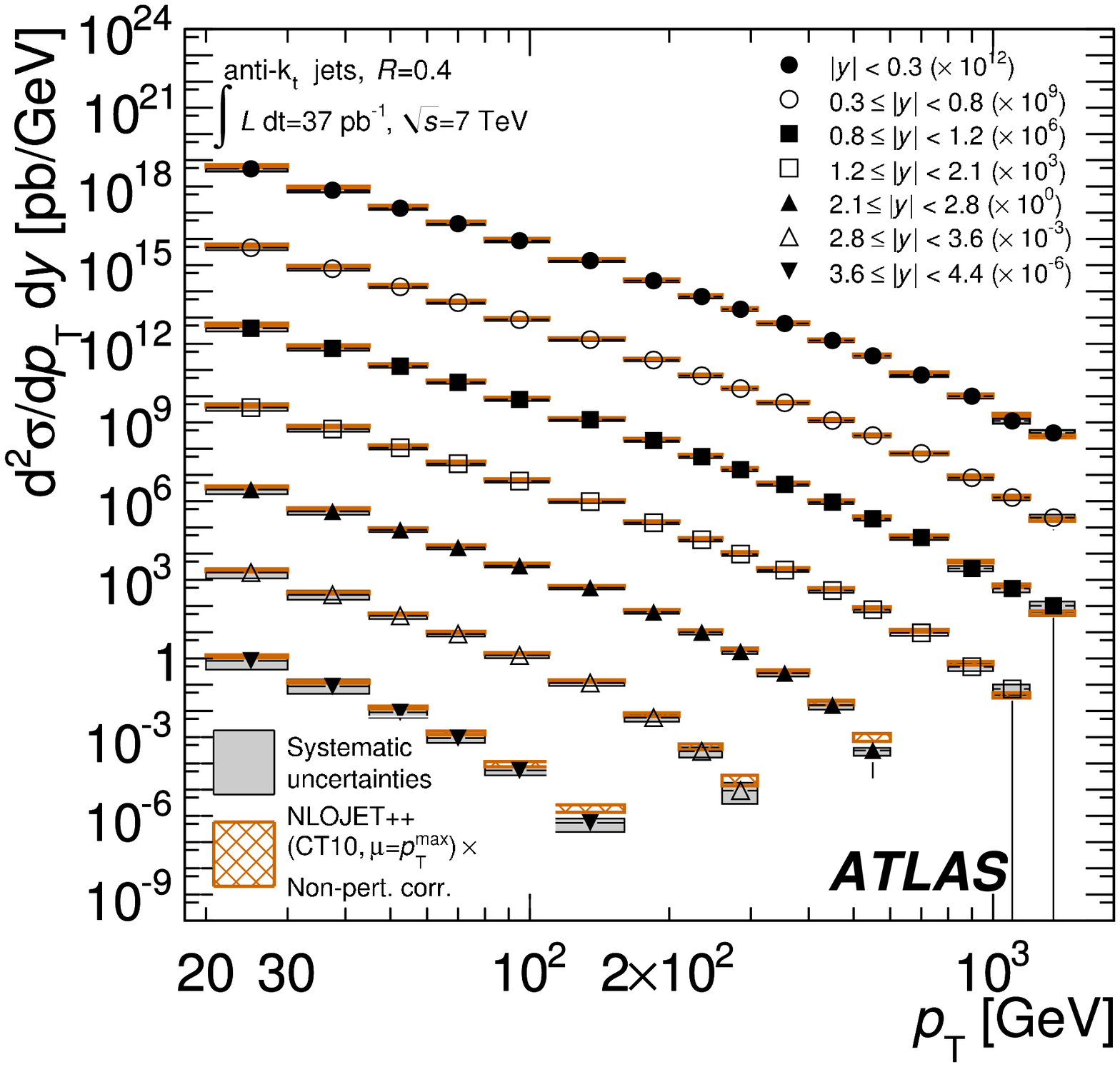}
}
\subfigure{
   \includegraphics[height=6.5cm]{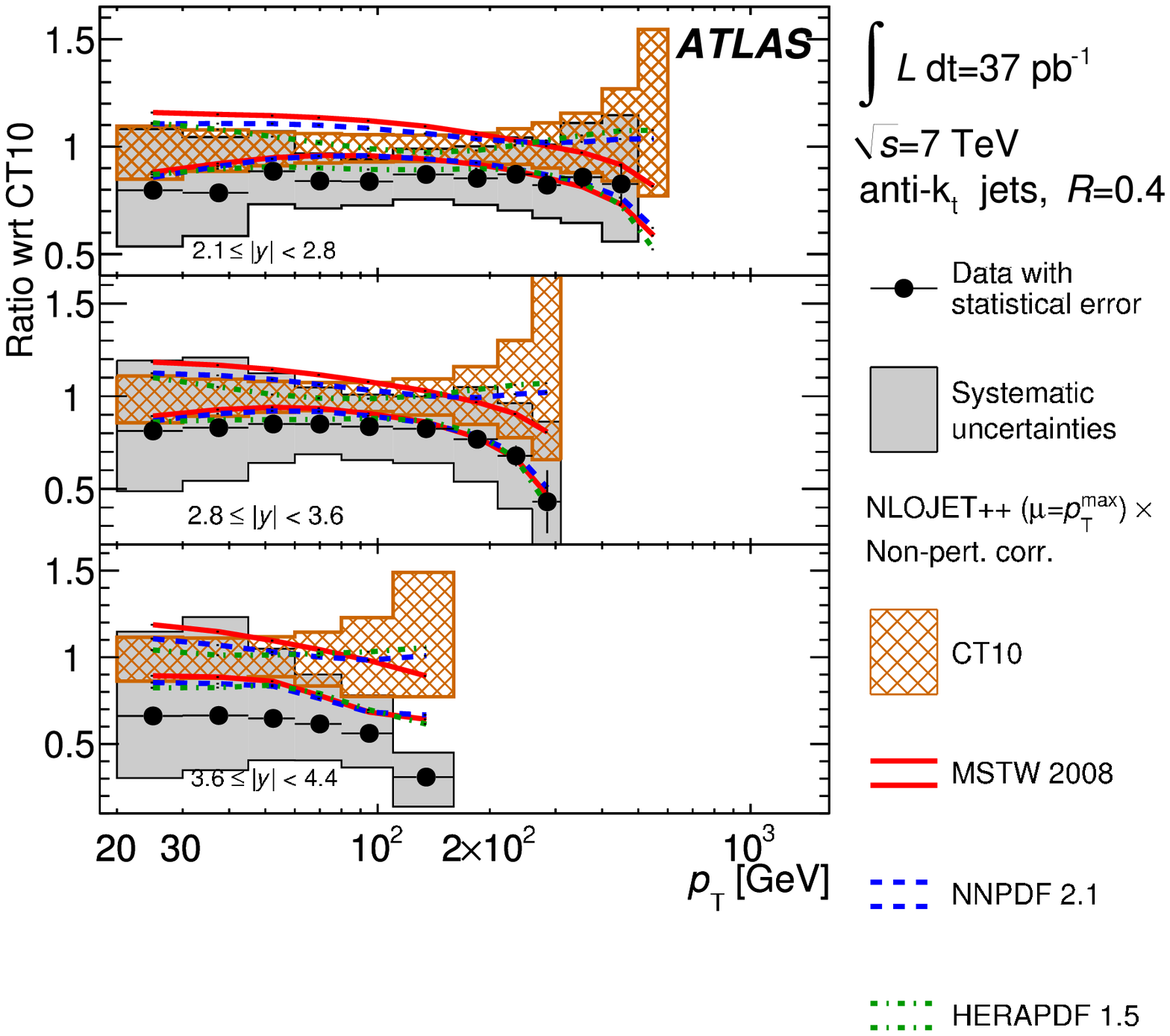}
}
\caption{Measurement of the inclusive jet cross section in the ATLAS detector.\protect\cite{Aad:2011fc}
A slight tension is observed between data and theory at high $p_{T}$.}
\label{fig:jetinc}
\end{figure}

The inclusive jet cross section measures the production rate of jets as a function of both transverse momentum ($p_{T}$) and rapidity ($y$).
In 2010 jets were measured with 20~GeV~$<p_{T}<$~1550~GeV out to rapidities of $|y| = 4.4$ using 37 pb$^{-1}$ of integrated luminosity at ATLAS,\cite{Aad:2011fc}\ with similar results seen in CMS.\cite{CMS:2011ab}
In 2011 the data sample of 4.8 fb$^{-1}$ has extended the reach to a $p_{T}$ of almost 2 TeV in CMS.\cite{CMS:2012dj}
The large reach of this basic observable offers a powerful test of the Standard Model over many orders of magnitude.
A next-to-leading order (NLO) calculation is performed using NLOJET++ \cite{Catani:1996vz}, with non-perturbative corrections applied to account for hadronisation and underlying event.
Monte Carlo events are also generated with POWHEG BOX,\cite{Alioli:2010xa} producing NLO matrix elements with parton showering which are then interfaced to PYTHIA or HERWIG for hadronisation.
There is generally good agreement seen between acceptance corrected measurement and theory, as shown in Figure \ref{fig:jetinc}.
At high $p_{T}$, especially for large values of $y$, a tension is observed with theory over estimating data.

\begin{figure}
\centering
\includegraphics[height=5.0cm]{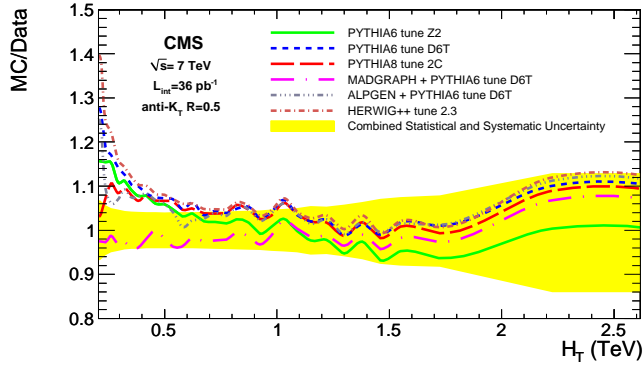}
\caption{Ratio of the cross section of events in CMS with $N \ge 3$ jets to $N \ge 2$ jets, where only jets with $p_{T} > 50$~GeV and $|y| < 2.5$ are considered. The ratio is shown as a function of $H_{T}$.\protect\cite{Chatrchyan:2011wn}}
\label{fig:threetworatio}
\end{figure}

Ratios of jet measurements are powerful because many systematics (jet energy calibration uncertainty and luminosity for example) either paritally or fully cancel when the ratio is taken.
Taking the ratio of events with $N \ge 3$ jets to $N \ge 2$ jets is an interesting probe of NLO effects.~\cite{Chatrchyan:2011wn}
Measured  as a function of the scalar sum of jets $p_{T}$, $H_{T} = \Sigma~jet~p_{T}$ where the sum is extended to all jets with $p_{T}>50$~GeV and $|y|<2.5$, Figure \ref{fig:threetworatio} shows that for $H_{T} > 500$~GeV a variety of MC predicts the data well.

\begin{figure}
\centering
\subfigure[Ratio of the inclusive dijet cross section to the exclusive dijet cross section in CMS.\protect\cite{Chatrchyan:2012pb}]{
   \includegraphics[height=6.0cm]{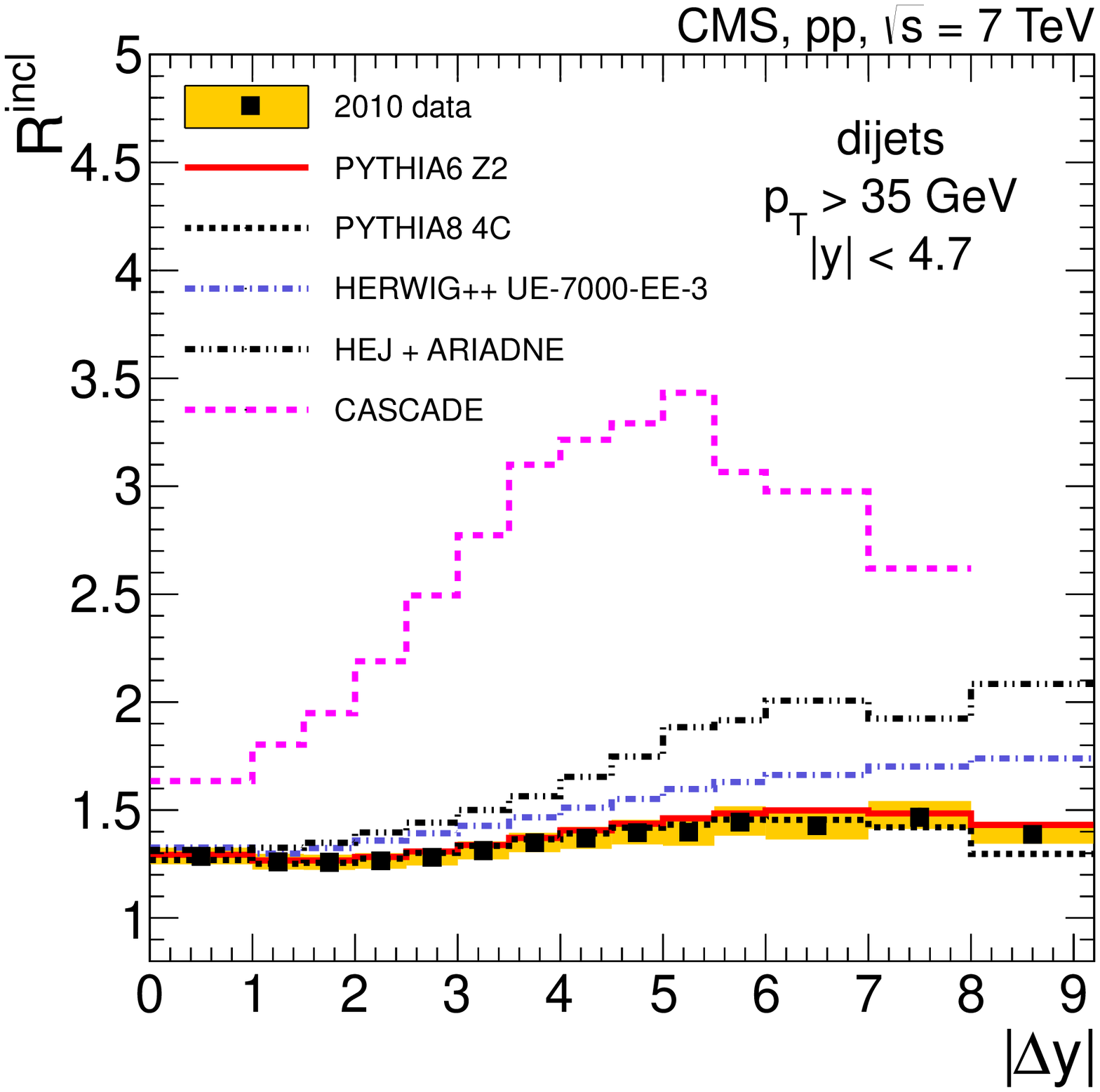}
   \label{fig:incexc}
} \hspace{0.4cm}
\subfigure[Fraction of jets containing a $D^{* \pm}$ meson in ATLAS.\protect\cite{Aad:2011td}]{
   \includegraphics[height=6.0cm]{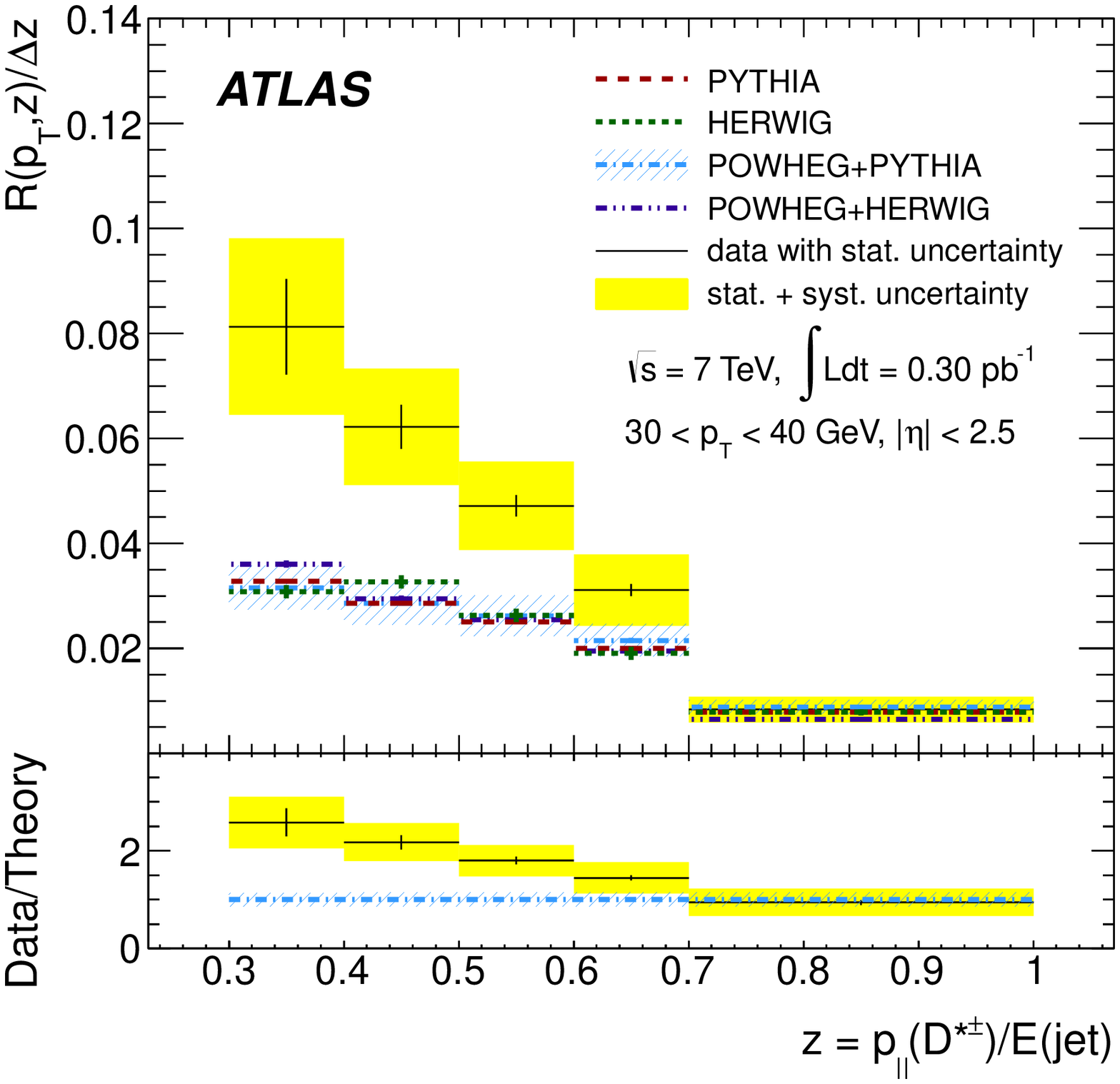}
   \label{fig:dstar}
}
\label{fig:incexcdstar}
\caption{Two ratio measurements from ATLAS and CMS.}
\end{figure}

The ratio of the inclusive dijet cross section (considering all combinations of $N \ge 2$ jets in an event) to the exclusive dijet cross section (only consider events with exactly $N = 2$ jets) is sensitive to the resummation of large $log(1/x)$ terms (BFKL evolution).
All jets with $p_{T} > 35$~GeV and $|y| < 4.7$ are considered, with the ratio plotted as a function of absolute rapidity separation $|\Delta y|$ between jet pairs.~\cite{Chatrchyan:2012pb}
As seen in Figure \ref{fig:incexc} PYTHIA gives the best agreement to data.

Heavy flavour at the LHC is important for understanding backgrounds in searches for the Higgs boson and/or super-symmetric particles, as well as providing a check of the hadronisation description in MC.
Figure \ref{fig:dstar} shows the ratio of jets containing a $D^{* \pm}$ meson to all jets as a function of $z$, the $D^{* \pm}$ momentum along the jet axis divided by the jet energy.~\cite{Aad:2011td}
For this low $p_{T}$ slice the agreement between data and MC is poor.
At low $p_{T}$, $D^{* \pm}$ originate mostly from c-hadrons showing that c-fragmentation in jets is not well modeled.

\begin{figure}
\centering
\subfigure[$b\bar{b}$ dijet cross section as a function of dijet mass in ATLAS.]{
   \includegraphics[height=6.0cm]{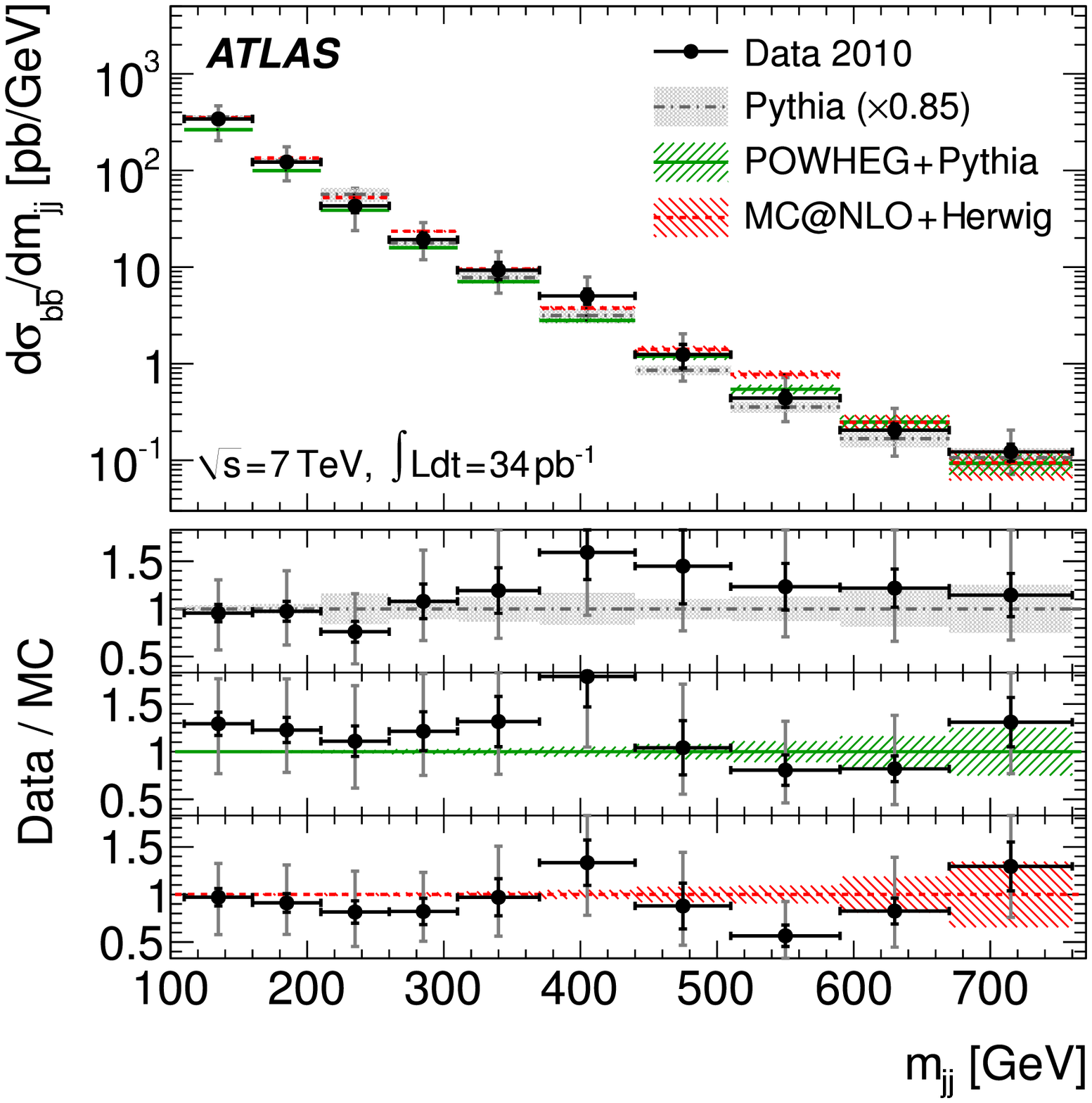}
   \label{fig:bjet:mass}
} \hspace{0.4cm}
\subfigure[$b\bar{b}$ dijet cross section as a function of $\Delta\phi$ in ATLAS.]{
   \includegraphics[height=6.0cm]{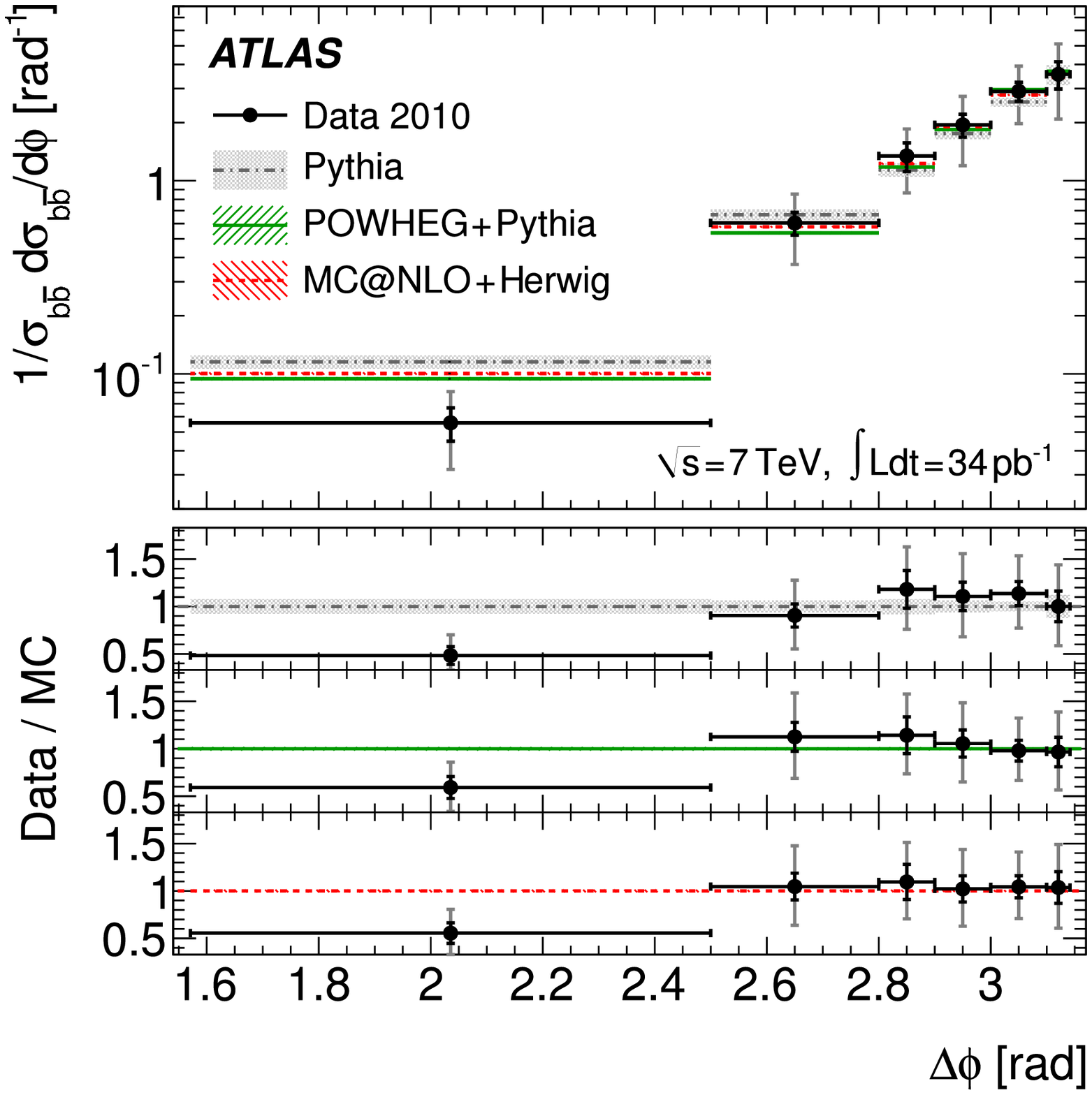}
   \label{fig:bjet:dphi}
}
\label{fig:bjet}
\caption{Results on jets produced from $b$-hadrons.\protect\cite{ATLAS:2011ac}}
\end{figure}

The measurement of the dijet cross section for jets from b-hadrons tests the production and hadronization of b-quarks.~\cite{ATLAS:2011ac}
Figure \ref{fig:bjet:mass} shows the dijet mass cross section from $b\bar{b}$ pairs, where theory is seen to describe data well.
For dijet systems which radiate a gluon, the azimuthal angle $\Delta\phi$ between them will be reduced.
Figure \ref{fig:bjet:dphi} shows that while back-to-back systems (larger $\Delta\phi$) are well described, as $\Delta\phi$ decreases both POWHEG+Pythia and MC@NLO+Herwig begin to over estimate the data.

\begin{figure}
\centering
\subfigure{
   \includegraphics[height=6.0cm]{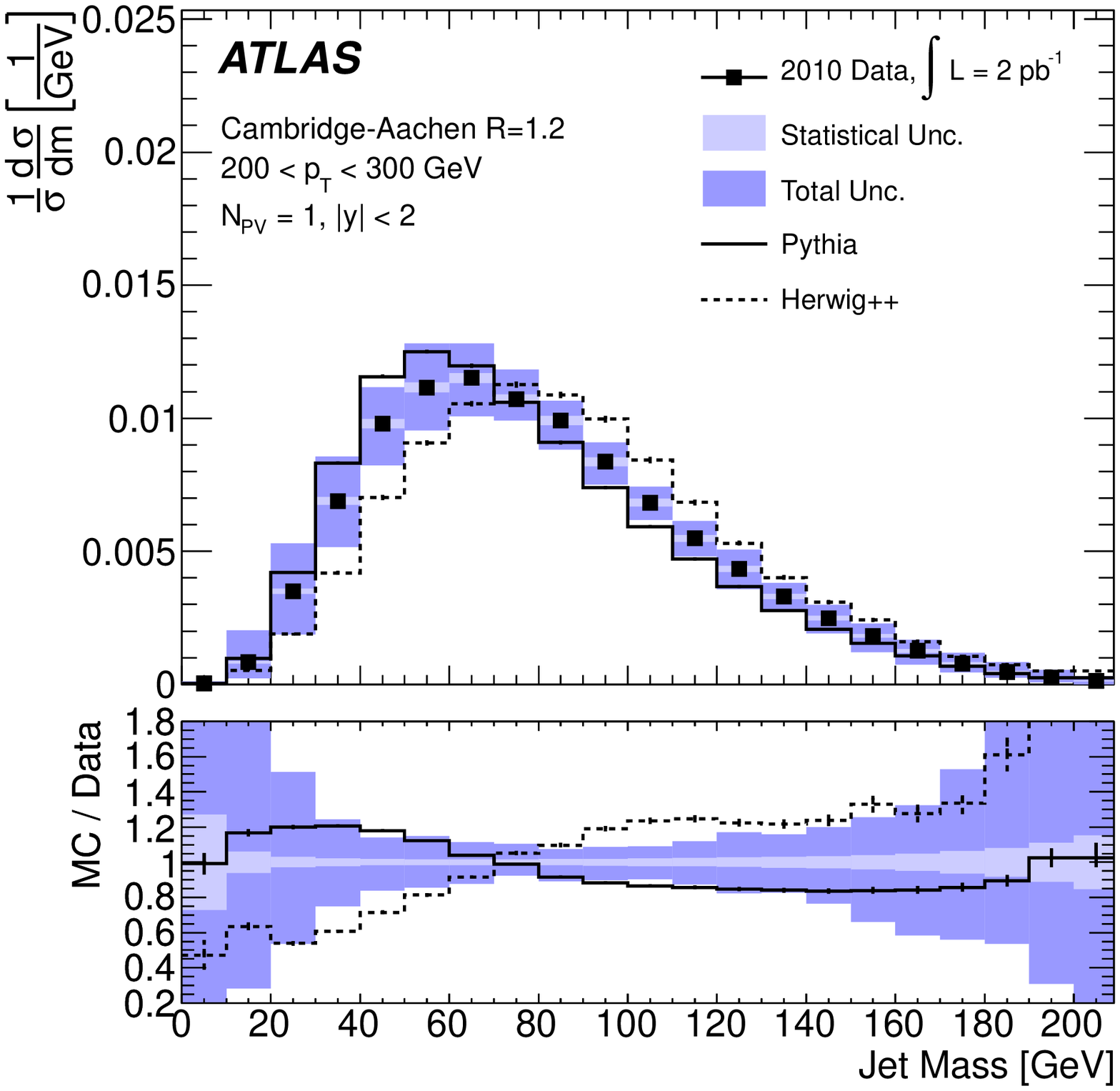}
} \hspace{0.4cm}
\subfigure{
   \includegraphics[height=6.0cm]{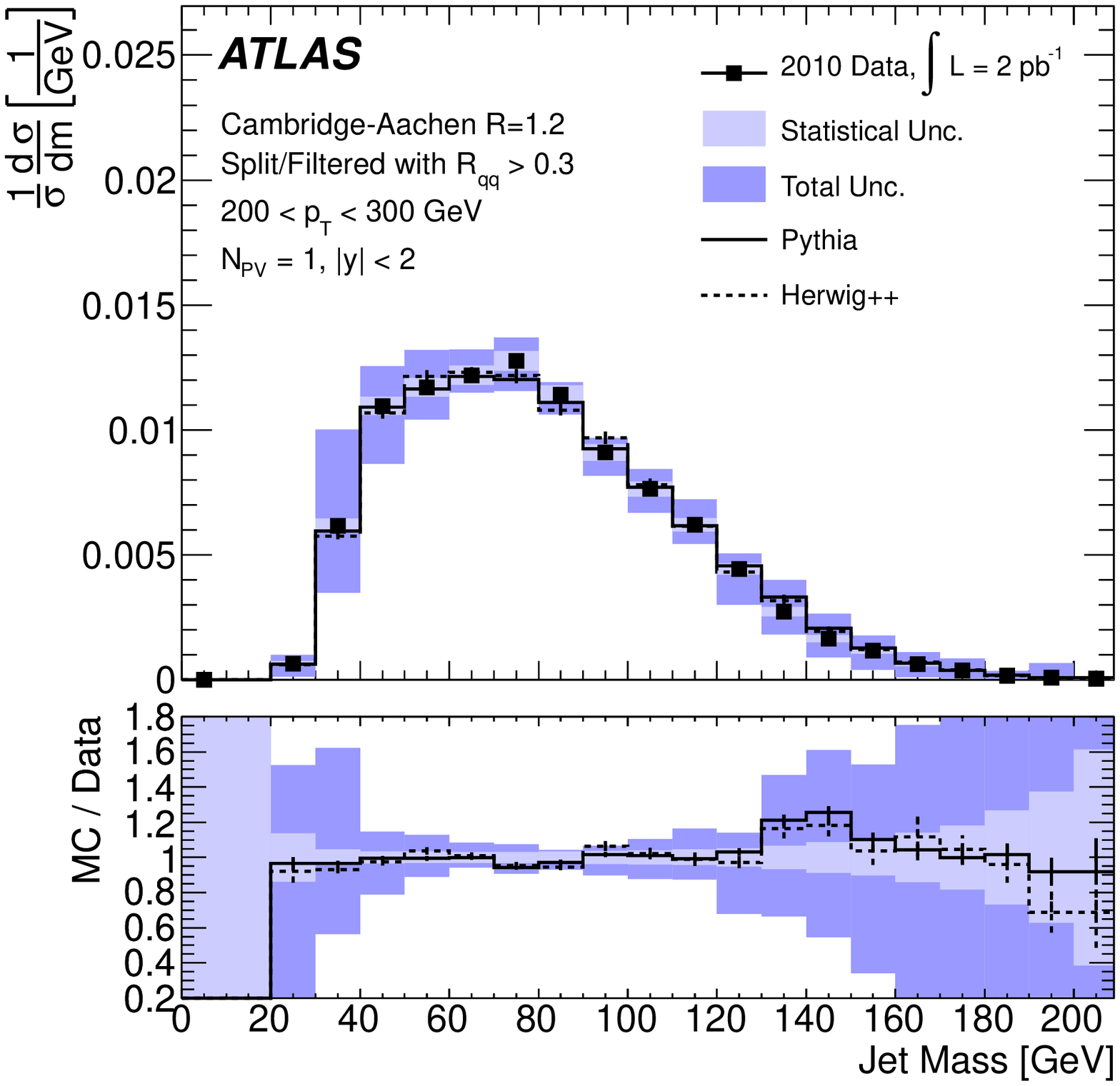}
}
\caption{Results from ATLAS on jet substructure for Cambridge-Aachen $R=1.2$ jets, before and after application of splitting/filtering.\protect\cite{ATLAS:2012am}}
\label{fig:sub}
\end{figure}

Jet substructure is useful for identifying hadronic decays of boosted heavy particles.
Splitting/filtering using Cambridge-Aachen $R=1.2$ jets is one example which undoes the clustering procedure until a large mass drop is observed.
This type of technique is robust against the effects of multiple proton-proton interactions in a single bunch crossing.
Figure \ref{fig:sub} shows the improved agreement between data and MC after splitting/filtering has been performed,\cite{ATLAS:2012am} giving confidence in the MC hadronisation description for substructure studies.

\section*{References}
\begin{thebibliography}{99}

\bibitem{ATLAS:det}
  ATLAS Collaboration,
  JINST {\bf 3} (2008) S08003.

\bibitem{CMS:det}
  CMS Collaboration,
  JINST {\bf 3} (2008) S08004.

\bibitem{Chatrchyan:2011je}
  CMS Collaboration,
  JINST {\bf 6} (2011) 11002.

\bibitem{Aad:2011he}
  ATLAS Collaboration,
  arXiv:1112.6426 [hep-ex].

\bibitem{Cacciari:2008gp}
  M.~Cacciari, G.~P.~Salam and G.~Soyez,
  JHEP {\bf 0804} (2008) 063.

\bibitem{Dokshitzer:1997in}
  Y.~L.~Dokshitzer, G.~D.~Leder, S.~Moretti and B.~R.~Webber,
  JHEP {\bf 9708} (1997) 001.

\bibitem{Aad:2011fc}
  ATLAS Collaboration,
  Phys.\ Rev.\ D {\bf 86} (2012) 014022.

\bibitem{CMS:2011ab}
  CMS Collaboration,
  Phys.\ Rev.\ Lett.\  {\bf 107} (2011) 132001

\bibitem{CMS:2012dj}
  CMS Collaboration,
  CMS-PAS-QCD-11-004 (2012),
  \url{https://cdsweb.cern.ch/record/1431022}.

\bibitem{Catani:1996vz}
  S.~Catani and M.~H.~Seymour,
  Nucl.\ Phys.\ B {\bf 485} (1997) 291
   [Erratum-ibid.\ B {\bf 510} (1998) 503].

\bibitem{Alioli:2010xa}
  S.~Alioli, K.~Hamilton, P.~Nason, C.~Oleari and E.~Re,
  JHEP {\bf 1104} (2011) 081.

\bibitem{Chatrchyan:2011wn}
  CMS Collaboration,
  Phys.\ Lett.\ B {\bf 702} (2011) 336.

\bibitem{Chatrchyan:2012pb}
  CMS Collaboration,
  arXiv:1204.0696 [hep-ex].

\bibitem{Aad:2011td}
  ATLAS Collaboration,
  Phys.\ Rev.\ D {\bf 85} (2012) 052005.

\bibitem{ATLAS:2011ac}
  ATLAS Collaboration,
  Eur.\ Phys.\ J.\ C {\bf 71} (2011) 1846.

\bibitem{ATLAS:2012am}
  ATLAS Collaboration,
  JHEP {\bf 1205} (2012) 128.
  
\end{thebibliography}
\end{document}